\documentclass[12pt,a4paper]{article}

\usepackage{scicite}

\usepackage{times}
\usepackage{amsmath}
\usepackage{amsfonts}
\usepackage{latexsym}
\usepackage{graphicx}
\usepackage{algorithm}
\usepackage{algorithmic}
\usepackage[margin=2.0cm]{geometry}
\usepackage{setspace}

\newenvironment{sciabstract}{%
\begin{quote} \bf}
{\end{quote}}

\newcounter{lastnote}

\title{Revealing spatial variability structures of geostatistical functional data via Dynamic Clustering}

\author
{Elvira Romano,$^{\ast}$ Antonio Balzanella, Rosanna Verde\\
\\
\normalsize{Department of Studi Europei e Mediterranei,}\\
\normalsize{Second University of Naples,  Via del Setificio 15, 81100  Caserta}\\
\\
\normalsize{$^\ast$To whom correspondence should be addressed; E-mail:  elvira.romano@unina2.it}
}

\date{}

\begin{document} 

\baselineskip24pt


\maketitle


\begin{sciabstract}

In several environmental applications data are functions of time, essentially continuous, observed and recorded discretely, and spatially correlated. Most of the methods for analyzing such data are extensions of spatial statistical tools which deal with spatially dependent functional data.
In such framework, this paper introduces a new clustering method. The main features are that it finds groups of functions that are similar to each other in terms of their spatial functional variability and that it locates a set of centers which summarize the spatial functional variability of each cluster. The method optimizes, through an iterative algorithm, a best fit criterion between the partition of the curves and the representative element of the clusters, assumed to be a variogram function.
The performance of the proposed clustering method was evaluated by studying the results obtained through the application on simulated and real datasets.
 
\end{sciabstract}


\section*{Introduction}

Spatial interdependence of phenomena  is a common feature of many environmental applications such as oceanography, geochemistry, geometallurgy, geography, forestry, environmental control, landscape ecology, soil science, and agriculture.
For instance, in daily patterns of geophysical and environmental phenomena where data (from temperature to sound) are instantaneously recorded over large areas, explanatory variables are functions of time, essentially continuous, observed and recorded discretely, and spatially correlated.

In the last years, the analysis of such data has been performed by Spatial Functional Data Analysis (SFDA) (Delicado et al. (2010)), a new branch of Functional Data Analysis (Ramsay, Silverman (2005)).  Most of the contributions in this framework are extensions of spatial statistical tools for functional data.

This paper focuses on clustering spatially related curves.


To the authors knowledge, existent clustering strategies for spatially dependent functional data are very limited.  
The approaches refer to the following main methods: hierarchical, dynamic, clusterwise and  model-based.
The hierarchical group of methods, (Giraldo et al. (2009)) is based on spatial weighted dissimilarity measures between curves. 
These are extensions to the functional framework of the approaches proposed for geostatistical data, where the norm  between curves is replaced by a weighted norm among the geo-referenced functions.
In particular, two alternatives are proposed for univariate and multivariate context, respectively. 
In the univariate framework, the weights correspond to the variogram values computed for the distance between the sites. In the multivariate framework, a dimensionality reduction is performed using a Principal Component Analysis technique for functional data  (Dauxois et al. (1982)) with the variogram values, computed on the first principal component, used as weights. 
The main characteristic of these approaches is in considering the spatial dependence among different kinds of functional data and in defining spatially weighted distances measures.

Alternatively to these approaches, with the aim of obtaining a partition of spatial functional data and a suitable representation for each cluster, the same authors proposed dynamic (Romano et al. (2010)) and clusterwise methods (Romano, Verde (2009)). 
The first, aims at classifying spatially dependent functional data and achieving a kriging spatio-functional model prototype  for each cluster by minimizing the spatial variability measure among the curves in each cluster.

In the ordinary kriging for functional data, the problem is to obtain an estimated curve in an unsampled location.
This proposed method gets not only a prediction of the curve but also a best representative location.     
In this sense, the location is a parameter to estimate and the objective function may have several local minima corresponding to different local kriging. The method proposes to solve this problem by evaluating local kriging on unsampled locations of a regular spatial grid in order to obtain the best representative predictor for each cluster.  This approach is based on the definition of a grid of sites in order to obtain the best representative function.
In a different manner and for several functional data, the clusterwise linear regression approach attempts to discover spatial functional linear regression models with two functional predictors, an interaction term, and spatially correlated residuals. This approach can establish a spatial organization in relation to the interaction among different functional data.
The algorithm is a  k-means clustering with a criterion based on the minimization of the squared
residuals instead of the classical within cluster dispersion.

A further approach is a model-based method for clustering
multiple curves or functionals under spatial dependence specified by a set of unknown parameters (Jiang, Serban (2010)). The functionals are decomposed using a
semi-parametric model, with fixed and random effects.  The fixed effects account for the large-scale
clustering association and the random effects account for the small scale spatial dependence
variability. 
Although the clustering algorithm is one of the first endeavors in
handling densely sampled space domains using rigorous statistical modeling, it presents several computational difficulties in applying 
the estimation algorithm  to a large number of spatial units.


The method proposed in this paper, belongs to the dynamic clustering approaches (Diday (1971)).
The current interest is motivated by a wide number of environmental applications where understanding the spatial relation among curves in an area is an important source of information for making a prediction regarding an unknown point of the space. 
The main idea is to provide  a summary of the set of curves spatially correlated by a prototype-based clustering approach. 
With this aim the proposed method uses a Dynamic Clustering approach to optimize a best fit criterion between the partition and the representative element of the clusters, assumed to be a variogram function. \footnote{A preliminary version of this paper appears in (Romano et al. (2010))}
According to this procedure, clusters are groups of functions that are similar to each other in terms of their spatial functional variability.
The central issue in the procedure consists in taking into account
the spatial dependence of georeferenced functional data. For most environmental applications, the spatial process is considered to be stationary and isotropic, and a wide area of the space is modeled
with a single variogram model. In practice, however, many spatial functional data cannot be modeled accurately with the same variogram model. Recognizing this, the scope is to propose a clustering method that clusters the geo-referenced curves into groups and associates a variogram function to each of them.  

The rest of this paper is organized as follows.  Section $1$ introduces the concept of spatial functional data and the measures for studying their spatial relation. Section $2$ shows the proposed method. Section $3$ illustrates the method on synthetic and real datasets.

\section{Spatial variability measure for geostatistical functional data}

Spatially dependent functional data may be defined as the data for which the measurements on each observation that is a curve are
part of a single underlying continuous spatial functional process defined as
\begin{equation}
\left\{\chi_{s}:\ s\in D\subseteq R^{d}\right\}
\end{equation}
where $s$ is a generic data location in the $d-$dimensional Euclidean space ($d$ is usually equal to $2$), the set $D\subseteq R^{d}$ can be fixed or random, and $\chi_{s}$ are functional random variables, defined as random elements taking values in an infinite dimensional space.
The nature of the set $D$ allows the classification of Spatial Functional Data. Following (Delicado et al. (2010)) these can be distinguished in geostatistical functional data, functional marked point patterns and functional areal data. 

The paper focuses on geostatistical functional data, where samples of functions are observed in different sites of a region (spatially correlated
functional data).

Let $\left\{\mathbf{\chi_{s}}(t): t \in T, s \in D\subset R^{d}\right\}$ be a random field where the set $ D\subset R^{d}$ is a fixed subset of $R^{d}$ with positive volume. $\mathbf{\chi_{s}}$ is a functional variable defined on some compact set $T$ of $R$ for any $s\in D$. 

It is assumed to observe a sample of curves $\left(\chi_{s_{1}}(t),\ldots, \chi_{s_{i}}(t), \ldots, \chi_{s_{n}}(t)\right)$ for $t \in T$ where $s_{i}$ is a generic data location in the $d$-dimensional Euclidean space.

For each $t$, the random process is assumed to be second order stationary and isotropic: that is, the mean and variance functions are constant and the covariance depends only on the distance between sampling sites.
Formally:
$\mathbb{E}(\mathbf{\chi_{s}}(t))=m(t)$, for all $t\in T,\:s\in D$, $\mathbb{V}(\mathbf{\chi_{s}}(t))=\sigma^{2}(t)$, for all  $t\in T,\:s\in D$, and
$\mathbb{C}ov(\chi_{s_{i}}(t),\chi_{s_{j}}(t))=C(h,t)$ where $h_{ij}=\left\|s_{i}-s_{j}\right\|$ and all $s_{i}, s_{j}\in D$

This implies that a variogram function for functional data $\gamma(h,t)$  exists, also called trace-variogram function (Giraldo et al. (2009)), such that

\begin{equation}
\gamma(h,t)=\gamma_{s_is_j}(t)=\frac{1}{2}\mathbb{V}(\chi_{s_{i}}(t)-\chi_{s_{j}}(t))=\frac{1}{2}\mathbb{E}\left[\chi_{s_{i}}(t)-\chi_{s_{j}}(t)\right]^{2}
\end{equation}
where $h=\left\|s_{i}-s_{j}\right\|$ and all $s_{i},s_{j}\in D$.

By using Fubini's theorem, the previous becomes $\gamma(h)=\int_{T}\gamma_{s_i s_j}(t) dt$ for $\left\|s_{i}- s_{j}\right\|=h$.
This variogram function can be estimated by the classical method of the moments by means of:
\begin{equation}
\label{varfunctionaldata}
\hat{\gamma}(h)=\frac{1}{2\left|N(h)\right|}\sum_{i,j\in N(h)}\int_{T}\left(\chi_{s_{i}}(t)- \chi_{s_{j}}(t)\right)^{2}dt
\end{equation}
where $N(h)=\left\{\left(s_{i}; s_{j}\right)\!:\!\left\|s_{i}- s_{j}\right\|=h\right\}$ for regular spaced data and $\left|N(h)\right|$ is the number of distinct elements in $N(h)$. 

When data are irregularly spaced, $N(h)=\left\{\left(s_{i}; s_{j}\right)\!:\!\left\|s_{i}- s_{j}\right\| \in\left(h-\epsilon, h+\epsilon\right)\right\}$ with $\epsilon\geq0$ being a small value.

The estimation of the empirical variogram for functional data using (\ref{varfunctionaldata}) involves the computation of integrals that can be simplified by considering that the functions are expanded in terms of some basis functions
\begin{equation}
\label{base}
\chi_{s_{i}}(t)=\sum^{Z}_{l=1}a_{il}B_{l}(t)={{\textbf{a}}_{i}}^{T}\textbf{B}(t),\ i=1,\ldots, n
\end{equation}
where $\textbf{a}_{i}$ is the vector of the basis coefficients for the $\chi_{s_{i}}$,  then the coefficients of the curves can be consequently organized in a matrix as follows:

\[ A=\left( \begin{array}{cccc}
a_{1,1} & a_{1,2} &\ldots & a_{1,Z}\\
a_{2,1} & a_{2,2}& \ldots & a_{2,Z}\\
\vdots & \ddots& \ldots & \ldots  \\
a_{n,1} & a_{n,2} &\ldots & a_{2,Z} \end{array} \right)_{n\times Z}
\]

Thus, the empirical variogram function for functional data can be obtained by considering:  

\begin{eqnarray*}
\int_{T}\left(\chi_{s_{i}}(t)- \chi_{s_{j}}(t)\right)^{2}dt & = &   \int_{T}\left({\textbf{{a}}_{i}}^{T}\textbf{B}(t)-{{\textbf{a}}_{j}}^{T}\textbf{B}(t)\right)^{2}dt = \\ & & =\int_{T}\left({{\textbf{{a}}_{i}}-{\textbf{{a}}_{j}}}\right)^{T}\textbf{B}(t)^{2}dt = \\ & & =\left({{\textbf{{a}}_{i}}-{\textbf{{a}}_{j}}}\right)^{T}\left(\int_{T}\textbf{B}(t){\textbf{B}(t)}^{T}dt\right)\left({{\textbf{{a}}_{i}}-{\textbf{{a}}_{j}}}\right)^{T}=\\ & &
=\left({{\textbf{{a}}_{i}}-{\textbf{{a}}_{j}}}\right)^{T}\mathbf{W}\left({{\textbf{{a}}_{i}}-{\textbf{{a}}_{j}}}\right)^{T}
\end{eqnarray*}
where $\mathbf{W}=\int_{T}{\textbf{B}\left(t\right){\textbf{B}(t)}^{T}dt}$ is the Gram matrix that is the identity matrix for any orthonormal basis.   For other basis as B-Spline basis function, $W$ is computed by numerical integration. 
Thus the variogram is expressed by:
\[
\gamma(h)=\frac{1}{2\left|N(h)\right|}\sum_{i,j\in N(h)}\left[\left(\textbf{a}_{i}-\textbf{a}_{j}\right)^{T}\mathbf{W}\left(\textbf{a}_{i}-\textbf{a}_{j}\right)\right]\;\forall i,j\ |\ \left\|s_{i}-s_{j}\right\|=h
\]

The empirical variograms cannot be computed at every lag distance $h$, and due to variation in the estimation, it is not ensured that it is a valid variogram. 

In applied geostatistics, the empirical variograms are thus approximated (by ordinary least squares (OLS) or weighted least squares (WLS)) by model functions, ensuring validity (Chiles, Delfiner (1999)). Some widely used models include: Spherical, Gaussian, exponential, or Mathern (Cressie (1993)).
The variogram, as defined before, is used to describe the spatial variability among functional data across an entire spatial domain. In this case, all possible location pairs are considered.

However, this spatial variability may be strongly influenced by an unusual or changing behavior within this wide area. 
For instance, in climatology, a sensor network is used to evaluate the temperature variability over an area. Some sensors could describe the characteristics of their surrounding sites with very different proportions, causing potentials for errors in the computation of spatial variability.

Thus, in order to describe these spatial variability substructures, this paper introduces the concept of the spatial variability components with regards to a specific location by defining a centered variogram for functional data.
 
Coherently with the above definition, given a curve $\chi_{s_{i}}(t)$, the centered variogram for functional data can be expressed by
\begin{equation}
\gamma^{s_i}(h,t)= \frac{1}{2}\mathbb{E}(\chi_{s_{i}}(t)-\chi_{s_{j}}(t))
\end{equation}
for each $s_j\neq s_i \in D$.
Similar to the variogram function, the centered variogram of the curve $\chi_{s_{i}}(t)$, as a function of the lag $h$, can be estimated through the method of moments:
\begin{equation}
\hat{\gamma}^{s_i}(h)=\frac{1}{2\left|N^{s_{i}}(h)\right|}\sum_{i,j\in {N^{s_{i}}(h)}}\int_{T}\left(\chi_{s_{i}}(t)- \chi_{s_{j}}(t)\right)^{2}dt 
\end{equation}

where $N^{s_{i}}(h)\subset N(h) = \left\{\left(s_{i}; s_{j}\right)\!:\!\left\|s_{i}- s_{j}\right\|=h\right\}$  and it is such that $\left|N(h)\right|=\sum_{i} \left|N^{s_{i}}(h)\right|$. 

Through straightforward algebraic operations, it is possible to show that the variogram function is a weighted average of centered variograms:
\begin{equation}
\hat{\gamma}(h)=\frac{1}{2\left|N(h)\right|}\sum_{i=1}^{n}\left(\frac{1}{2\left|N^{s_{i}}(h)\right|} \sum_{i,j\in N^{s_{i}}(h)}\int_{T}\left(\chi_{s_{i}}(t)- \chi_{s_{j}}(t)\right)^{2}dt   \right)2\left|N^{s_{i}}(h)\right|	
\end{equation}
thus:

\begin{equation}
\hat{\gamma}(h)= \frac{1}{2\left|N(h)\right|}\sum_{i=1}^{n} \hat{\gamma}^{s_i}(h) 2\left|N^{s_{i}}(h)\right|
\end{equation} 
     
It is worth noting that the estimation of the centered variogram can be expressed in the same manner in the functional setting. 



\section{Variogram-based Dynamic Clustering approach for spatially dependent functional data}
\label{subsec:2} 
A Dynamic Clustering Algorithm (DCA) (Celeux et al. (1988)) (Diday (1971)) is an unsupervised learning algorithm, which finds partitions a set of objects into internally dense and sparsely connected clusters. 
The main characteristic of the DCA is that it finds, simultaneously, the partition of data into a fixed number of clusters and a set of representative syntheses, named prototypes, obtained through the optimization of a fitting criterion. 
Formally, let $E$ be a set of $n$ objects. The Dynamic Clustering Algorithm finds a partition $P^*=(C_1,\ldots,C_k,\ldots,C_K)$
of $E$ in $K$ non empty clusters and a set of representative
prototypes $L^*=(G_1,\ldots,G_k,\ldots,G_K)$ for each $C_k$ cluster of
$P$ so that both $P^*$ and $L^*$ optimize the following criterion:

\begin{equation}\label{eq00}
  \Delta(P^*,L^*) = Min \, \{\Delta(P,L) \; / \; P \in P_K, L \in \Lambda_K\}
\end{equation}
with $P_K$ the set of all the $K$-cluster partitions of $E$ and
$\Lambda_K$  the representation space of the prototypes.
$\Delta(P,L)$ is a function, which measures how well the prototype $G_{k}$ represents the characteristics of objects of the cluster and it can usually be interpreted as an heterogeneity or a dissimilarity measure of goodness of fit between $G_{k}$ and $C_{k}$.

The definition of the algorithm is performed according to two main tasks:

\begin{itemize}
\item[-] {\it representation function} allowing to associate to
each partition $P \in P_K$ of the data in $K$ classes $C_k$ $(k=1,
\ldots, K$), a set of prototype $L= (G_1,\ldots,G_k,\ldots,G_K)$ of the representation space $\Lambda_K$

\item[-] {\it allocation function} allowing to assign to each
$G_{k} \in L$, a set of elements $C_k$.
\end{itemize}

The first choice concerns the representation structure $L$ for the
classes ${C_{1}, \ldots, C_{K}} \in P$. 

Let $\left\{\chi_{s_{1}}(t),\ldots, \chi_{s_{n}}(t)\right\}$ (with $t \in T$ and $s\in D$) be the sample of spatially located functional data. 
The proposed method aims at partitioning them into clusters in order to minimize, in each cluster, the spatial variability. 

Following this aim, the method optimizes a best fit criterion between the centered variogram function $\gamma^{s_{i}}_{k}(h)$ and a theoretical variogram function $\gamma_k^*(h)$ for each cluster as follows:

\begin{equation}
\label{eq01}
  \Delta(P,L) = \sum_{k=1}^K \sum_{\chi_{s_{i}}(t)\in C_k} ({\gamma}^{s_i}_{k}(h) - \gamma_k^*(h))^2
\end{equation}

where ${\gamma}^{s_i}_{k}$ is the centered variogram,  which describes the spatial dependence between a curve $\chi_{s_{i}}(t)$ at the site $s_{i}$ and all the other curves $\chi_{s_{j}}(t)$ at different spatial lags $h$. This allows to evaluate the membership of a curve $\chi_{s_{i}}(t)$ to the spatial variability structure of an area.



As already mentioned, starting from a random initialization, the algorithm alternates
\textit{representation} and \textit{allocation} steps until it reaches the convergence to a stationary value of the criterion $\Delta(P,L)$.

In the \textit{representation}  step,  the theoretical variogram $\gamma^*_k(h)$ of the set of curves $\chi_{s_{i}}(t)\in C_k$, for each cluster $C_k$ is estimated. This involves the computation of the empirical variogram and its model fitting  by the Ordinary Least Square method. 

In the \textit{allocation} step, the function ${\gamma}^{s_i}_{k}$ is computed for each curve $\chi_{s_{i}}(t)$. Then a curve $\chi_{s_{i}}(t)$ is allocated to a cluster $C_k$ by evaluating its matching with the spatial variability structure of the clusters according to the following rule: 

\begin{equation}
\sum_{h<h^{*}\in\left[m_{k}; M_{k}\right]}({\gamma}^{s_{i}}_{k}(h) - \gamma_k^*(h))^2 \rho_{k}< \sum_{h<h^{*}\in\left[m_{k}; M_{k}\right]}({\gamma}^{s_{i}}_{k^{'}}(h) - {\gamma}_{k^{'}}^{*}(h))^{2}\rho_{k^{'}} \ \ \forall k\neq k^{'}
\end{equation}
where:
\begin{itemize}
	\item $\rho_{k}=\frac{\left|{N^{s_{i}}}_{k}\right|}{\left|N_{k}\right|}$ and $\rho_{k^{'}}=\frac{\left|N^{s_{i}}_{k^{'}}\right|}{\left|N_{k^{'}}\right|}$ are the weights computed respectively, considering, for a fixed $s_{i}$, the number of location pairs $N^{s_{i}}_{k}$, $N^{s_{i}}_{k^{'}}$ that are separated by a distance $h$ in a cluster $k$, and $k^{'}$.  

\item 
$m_{k}=\min_{k}h_{k}^{*}\:,M_{k}=\max_{k}h_{k}^{*}$ where $h_{k}^{*}$  is the spatial distance at which the variogram $\gamma_{k}^{*}$ for each cluster $k$ reaches its sill. 
\end{itemize}
The problem is that for each cluster, there are several values of $h_{k}^{*}\ \left(k=1,\ldots, K\right)$, due to the different spatial functional variability structures of the partition. 
According to the above allocation criterion, only one level $h^{*}$ is chosen such that for $h>h^{*}$, there is no spatial correlation. This rule facilitates the spatial aggregation process leading to a tendency to form regions of spatially correlated curves. Especially, $h^{*}$ is set in the range $\left[m_{k},M_{k}\right]$.

The consistency between the representation of the clusters and the allocation criterion guarantees the convergence of the criterion to a stationary minimum value (Celeux et al. (1988)). 

In the context of the proposed method, this is verified when:

\begin{equation}
	\gamma_k^*(h)= argmin \sum_{\chi_{s_{i}}(t)\in C_k} (\gamma^{s_i}_{k}(h) - \gamma_k^*(h))^2
\end{equation}

Thus, since the allocation of each curve $\chi_{s_{i}}(t)$ to a cluster $C_k$ is based on computing the squared Euclidean distance between $\gamma^{s_i}_{k}(h)$ and  $\gamma_k^*(h)$, since the variogram $\gamma_k^*(h)$ is the average of the functions $\gamma^{s_i}_{k}(h)$, then $\gamma_k^*(h)$ minimizes the spatial variability of each cluster.

\begin{algorithm}[htbp]
\caption{Dynamic Clustering Algorithm for geostatistical functional data}

\begin{algorithmic}
\label{algo:GUS}
\STATE \textit{Initialization}:

\STATE Start from a random partition $P=(C_1,\ldots,C_k,\ldots,C_K)$

\STATE \textit{Representation step}:
\FORALL {clusters $C_{k}$} 
\STATE Compute the prototype $\gamma^*_k(h)$ which optimizes the best fitting criterion:
\begin{equation*}
  \min{\sum_{\chi_{s_{i}}(t)\in C_k} (\gamma^{s_i}_{k}(h) - \gamma_k^*(h))^2}
\end{equation*}
\ENDFOR
\STATE \textit{Allocation step:}
\FORALL {$\chi_{s_{i}}(t)$ with $i=1,\ldots,n$}
\STATE find the cluster index $k$, for $h^{*}\in \left[m_{k}; M_{k}\right]$:
\STATE $\chi_{s_{i}}(t)\rightarrow C_k  \ \ \ if \ \ \sum_{h<h^{*}}({\gamma^{s_{i}}}_{k}(h) - \gamma_k^{*}(h))^2 \rho_{k}< \sum_{h<h^{*}}(\gamma^{s_i}_{k^{'}}(h) - \gamma_{k^{'}}^{*}(h))^{2}\rho_{k^{'}}\ \forall k\neq k^{'}$

\ENDFOR
\end{algorithmic}
\end{algorithm}

\section{Dealing with simulated and real data}
\label{application}

The performance of the proposed clustering method was evaluated by studying the results obtained through the application on simulated and real datasets.

\subsection{Test on Simulated data} 

First datasets are generated from a spatio functional random field with different spatial functional variability structure.

Specifically, given a sample of curves $\left(\chi_{s_{1}}(t),\ldots, \chi_{s_{i}}(t), \ldots, \chi_{s_{n}}(t)\right)$ for $t \in T$ where $s_{i}$ is a generic data location in the $d$-dimensional Euclidean space, and $\chi_{s_{i}}(t)$ is generated by a spatio-functional Gaussian random field. 

The primary scope is to test the performances of the procedure in detecting spatio functional variability structures.
Thus, it is considered a situation largely used in geostatistics, where the covariance between $\chi_{s_{i}}(t),\chi_{s_{j}}(t)$ is a stationary separable function of the form: 
\begin{equation}
C_{SEP}\left(h,u\right)=cov\left\{\chi_{s_{i}}(t),\chi_{s_{j}}(t)\right\}=C_{s}\left(h\right)C_{T}\left(u\right)
\end{equation}
where $C_{s}\left(h\right)$ and $C_{T}\left(u\right)$ are stationary, purely spatial and purely temporal covariance functions, respectively, defined on two generic locations $s_{i},s_{j}$  that are apart by $h=s_{i}-s_{j}$ with a time span $u=\left|t_{i}-t_{j}\right|$.
 
The simulation schema proposed by (Sun, Genton (2011)) is considered as reference.
In particular the spatial covariance function has the following form:
 \begin{equation}
C_{s}(h)=\left(1-\nu\right)\exp\left(-c\left|h\right|\right)+\nu{\delta}_{h=0}
\end{equation}
where $c>0$ controls the spatial correlation intensity, and $\nu \in \left(0,1\right]$ is the nugget effect; the  temporal covariance function is of the Cauchy type having the following form:
\begin{equation}
C_{T}\left(u\right)={\left(u+a\left|u\right|^{2\alpha}\right)}^{-1}
\end{equation}
where $\alpha\in\left(0, 1\right]$  controls the strength of the temporal correlation and $a > 0$ is the scale
parameter in time. 

Six datasets made by $n = 300$ curves located on a regularly spaced grid have been generated. The
following model is used:
\begin{equation}
\label{mod sim}
\chi_{s}(t)=\mu_{s}(t)+\epsilon_{s}(t)\ t\in T
\end{equation} 
with mean $\mu_{s}(t)=0$ and $\epsilon_{s}(t)$ is a Gaussian random field with zero mean and covariance function as defined above.
Each simulated dataset is made by curves belonging to three clusters $C_1, C_2, C_3$. Each cluster includes $100$ spatially adjacent curves generated according to the parameter sets in table
\ref{tab:ParametersForSimulatedDatasets}.

In each dataset and in each cluster there is no nugget effect ($\nu=0$); moreover, the other parameters are set to $a = 1$ and $\alpha=0,1$.

There are two basic scenarios  which are different in the values of standard deviation $\sigma$ used for generating the Gaussian random field of a cluster, so that the datasets $1,2,3$  belong to the first scenario, while the datasets $4,5,6$ belong to the second one.

The datasets of both scenarios are designed to get three different levels of spatial correlation intensity $c$.

\begin{table*}[h]
	\centering
		\begin{tabular}{|l|l|l|l|l|l|l|}
\hline
&\multicolumn{3}{|c|}{Values of $\sigma$}&\multicolumn{3}{|c|}{Values of $c$}\\
\hline
Dataset Id & $C_1$ & $C_2$ & $C_3$ & $C_1$ & $C_2$ & $C_3$\\
\hline
$1$ & $5$ &	$10$ &	$15$ &	$3$ &	$7$ &	$10$ \\
$2$ & $5$ &	$10$ &	$15$ &	$5$ &	$7$ &	$9$ \\
$3$ & $5$ &	$10$ &	$15$ &	$3$ &	$9$ &	$15$ \\
$4$ & $7$	& $10$ &	$13$ &	$3$ &	$7$ &	$10$ \\
$5$ & $7$ &	$10$ &	$13$ &	$5$ &	$7$ &	$9$\\
$6$ & $7$	& $10$ &	$13$ &	$3$ &	$9$ &	$15$\\
\hline
			
\end{tabular}

	\caption{Parameters for simulated datasets}
	\label{tab:ParametersForSimulatedDatasets}
\end{table*}

In order to evaluate the capability of the proposed method to discover the spatial variability structures in the data and the curves which concur to form them, the well known Rand Index (Rand (1971)) is used. This index, whose value is in the range $[0,1]$, allows the measurement of the degree of consensus between two partitions so that the value $0$ indicates that the two partitions do not agree on any pair of items while $1$ means that the partitions are exactly the same.

The test consists in computing the Rand Index between the true partition of data which emerges from the simulation schema and the partition given as output by the proposed clustering method. Since the latter depends on the initial random partitioning of data, the following table reports, for each dataset, the average Rand Index calculated on $100$ repetitions of the algorithm.

\begin{table*}[h]
	\centering
		\begin{tabular}{|l|l|}
\hline
Dataset Id & Average Rand Index \\
\hline
$1$ & $0.88$\\
$2$ & $0.87$\\
$3$ & $0.85$\\
$4$ & $0.84$\\
$5$ & $0.82$\\
$6$ & $0.79$\\
\hline
			
\end{tabular}

\caption{Rand Index value for each simulated dataset.}
\label{tab:RandIndex}
\end{table*}

The clustering results for the six datasets reflect the expectations based on the simulations. The RI appears to be high for all the simulated  datasets, especially for the first dataset, where the value is $0.88$. 
\begin{figure}[ht]
\includegraphics[scale=.50]{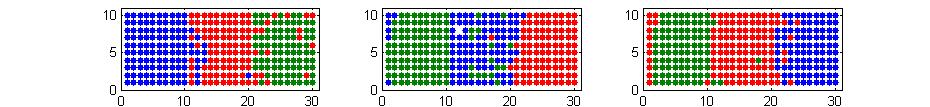}
%
%
\caption{Clustering results plotted on the spatial grid for the datasets $1,2,3$. The color of the dots identifies the cluster membership.}
\label{fig:1}       
\end{figure}

\begin{figure}[ht]
\includegraphics[scale=.50]{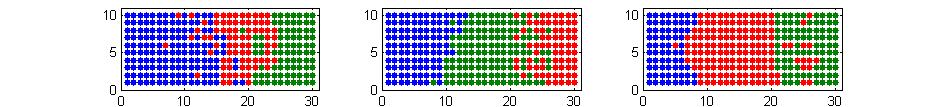}
%
%
\caption{Clustering results plotted on the spatial grid for the datasets $4,5,6$. The color of the dots identifies the cluster membership.}
\label{fig:2}       
\end{figure}

\begin{figure}[ht]
\center
\includegraphics[scale=.5]{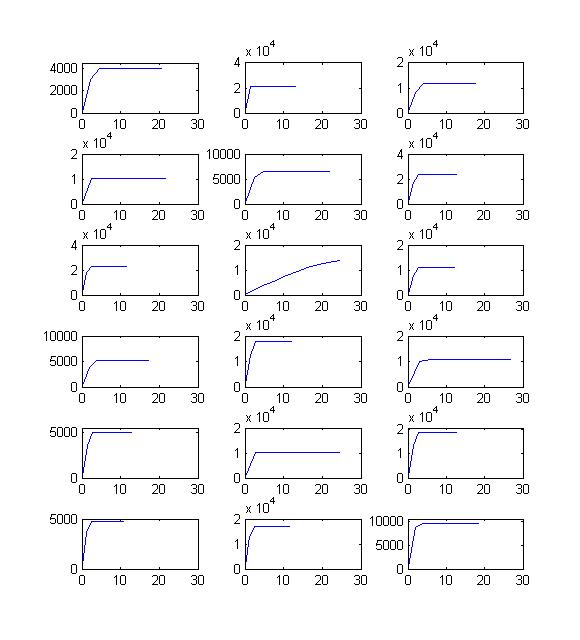}
%
%
\caption{Theoretical variogram functions for the simulated datasets}
\label{variograms}       
\end{figure}

The results are very interesting, since the clustering structures in data are discovered.
The good performance of the method is also highlighted by a graphic representation in Figure $1,2$, which plots the spatial locations of the three different clusters. Finally, Figure \ref{variograms} highlights the different variability structures through clusters prototypes.    

\subsection{Test on real data}

In order to evaluate the performance of the proposed strategy on real data, a dataset was provided by the Institute for Mathematics Applied to Geosciences \footnote{http://www.image.ucar.edu/Data/US.monthly.met/}.
The dataset reports the average monthly temperatures recorded by approximately 8000 stations located in the US, in the period 1895 to 1997.

Tests used data from $1993-1997$; thus for each station there is a time series made by a maximum of $60$ observations. Since for several stations there are no data in the considered period, the dataset is composed of $4500$ time series.

The first step of the analysis is to construct the set of functions expanded in terms of $B$-Spline Basis functions (\ref{base}).
An appropriate order of expansion $Z$ is chosen, taking into account that a large $Z$ causes overfitting and a too-small $Z$ may cause important aspects of the function to be missing of the estimated function (Ramsay, Silverman (2005)). They consider a procedure based on a classical non-parametric cross-validation analysis.
For each series, cubic splines are evaluated in order to produce a collection of smooth curves that is able to take into account the variability of the data.

The very large extension of the spatial region involved in the monitoring activity makes it difficult to apply
geostatistics methods based on the assumption of stationarity.  Since stationarity and isotropy are assumed in the strategy 
the spatial trend is removed in a first step of the analysis by using a functional regression model with functional response (smoothed temperature curves) and two scalar covariates (longitude and latitude coordinates in decimal degrees) (Giraldo et al. (2009)).

On these spatially located curves, it is evaluated the capability of the proposed strategy in discovering different variability structures and their associated spatial regions. 

In order to run the clustering algorithm, the following input parameters have to be set:

\begin{itemize}
	\item the number of clusters $K$
	\item the theoretical variogram model to fit the empirical one for each cluster
\end{itemize}

Since there is not any information on the true number of spatial variability structures, the algorithm is applied for $K= 2,\ldots,6$ and then $K$ is selected according to the maximum decreasing of the value of the optimized criterion $\Delta(P,L)$. For the tested dataset the best choice is $K=3$.

The theoretical variogram model is chosen evaluating several well known parametric models: Esponential, Spherical, Gaussian. The procedure is run for each model starting from the same initialization and then the fitting of each model to the data is evaluated, measuring the value of the criterion $\Delta(P,L)$. The results in Table \ref{VariogramModel} highlight that the best model is the exponential variogram thus, it is used on the tested dataset.

\begin{table*}[htbp]
	\centering

\begin{tabular}{p{3.0cm}p{3.0cm}}
\hline\noalign{\smallskip}
Trace-variogram model & $\Delta(P,L)$\\
Exponential & $2.9e^{+4}$ \\
Gaussian & $3.5e^{+4}$ \\
Spherical & $3.6e^{+4}$ \\

\noalign{\smallskip}\hline\noalign{\smallskip}

		\end{tabular}
	\caption{Criterion evaluation for several theoretical variogram models.}
	\label{VariogramModel}
\end{table*}

\begin{figure}[ht]
\center
\includegraphics[scale=.50]{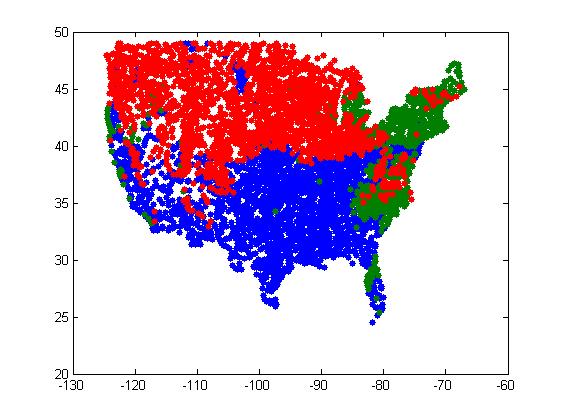}
%
%
\caption{Clusters plotted on the geographical map.}
\label{clusters}       
\end{figure}

Starting from the chosen input parameters, the algorithm run on the dataset, detects the spatial regions available in Fig. \ref{clusters}. The value of the optimized criterion is $\Delta(P,L)=2.9e^{+4}$; the number of iterations until convergence is $9$.

It is possible to note that the three discovered clusters split the studied area into three spatial regions, which include most of the east and west coasts, a northern area and a southern area.

\begin{figure}[ht]
\center
\includegraphics[scale=.30]{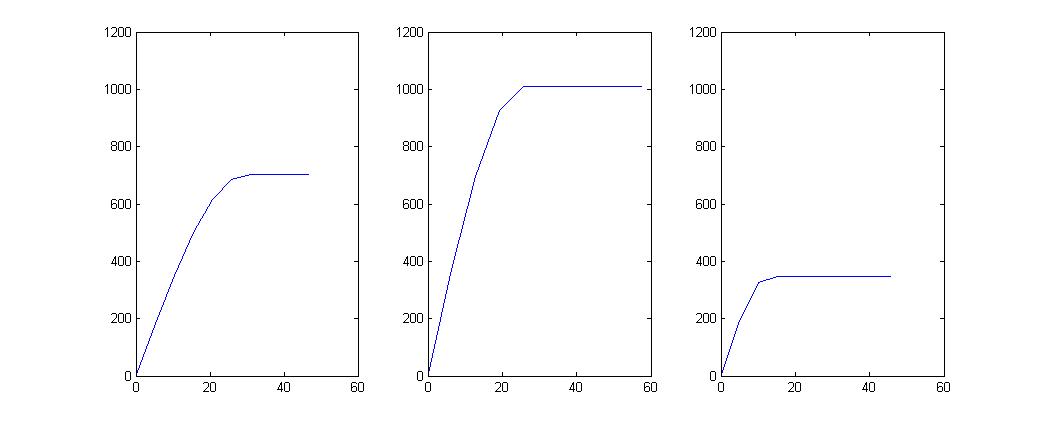}
%
%
\caption{Theoretical variogram models for the three discovered clusters.}
\label{variograms2}       
\end{figure}

These spatial regions are characterized by three different spatial variability structures as shown in Fig.\ref{variograms2}. 

It is possible to note that the variogram corresponding to the third cluster shows the lowest level of variance (sill); the second cluster presents a variogram with highest sill level. The range of the variograms is $29$ for the first cluster, $25$ for the second cluster and $16$ for the third one. 

Looking at the plots, it is possible to note that the variability in the first and second clusters rises at a lower rate when it is compared to the third cluster.

\section{Summary and conclusions}

This paper has introduced an exploratory strategy for geostatistical functional data.
  
It is a dynamic clustering method that partitions a set of geostatistical functional data into clusters that are homogeneuos in terms of spatial variability and that represents each cluster with a prototype variogram function. 

The approach is distinct from others since it discovers both the spatial partition of the data and the spatial variability
structures representative of each cluster. The spatial information is incorporated into the clustering process by considering the variogram as a measure of spatial
association, emphasizing the average spatial dependence among curves.

This strategy can represent a very interesting methodological proposal for analyzing georeferenced curves in which spatial dependence plays an important role in exploring the similarity among curves. 
As in classical geostatistics data analysis, it assumes that the process generating data is stationary and isotropic.
However, an alternative would be to consider an anisotropric process where the spatial dependence changes with the direction. 
In this case, it would be interesting to introduce a directional variogram model for functional data and demonstrate the main characteristics.

\section*{References}  
\begin{itemize}

\item Celeux, G. , Diday, E. , Govaert, G. , Lechevallier, Y. , Ralambondrainy, H. 1988. \emph{Classiffication Automatique des Donnees : Environnement Statistique et Informatique} - Dunod, Gauthier-Villards, Paris.

\item Chiles, J. P., Delfiner, P. 1999. \emph{Geostatististics, Modelling Spatial Uncertainty}. Wiley-Interscience.

\item Cressie, N. 1993. \emph{Statistics for spatial data}. Wiley Interscience.

\item Dauxois, J., Pousse, A., Romain, Y. 1982. Asymptotic theory for the principal component analysis of a vector random function: Some applications to statistical inference. \emph{Journal of Multivariate Analysis}, 12, 136-154. 

\item Delicado, P., Giraldo, R., Comas, C. and Mateu, J. 2010. Statistics for spatial functional data: some recent contributions. \emph{Environmetrics}, 21: 224–239.  

\item Diday, E. 1971. La methode des Nuees dynamiques. \emph{Revue de Statistique Appliquee}, 19, 2, 19-34.

\item Giraldo, R., Delicado, P., Comas, C., Mateu, J. 2009. Hierarchical clustering of spatially correlated functional data.
Technical Report. Available at: 

www.ciencias.unal.edu.co/unciencias/data-file/estadistica/RepInv12.pdf.

\item Giraldo, R., Delicado, P., Mateu, J. 2010. Ordinary kriging for function-valued spatial data. \emph{Journal of Environmental and Ecological Statistics}. Accepted for publication.

\item Jiang, H., Serban, N. 2010. Clustering Random Curves Under Spatial Interdependence: Classification of Service Accessibility. \emph{Technometrics}. 

\item Ramsay, J.E., Silverman, B.W. 2005. \emph{Functional Data Analysis} (Second ed.).Springer.

\item Rand,  W.M. 1971. Objective criteria for the evaluation of clustering methods. \emph{Journal of the American Statistical Association}.  Vol. 66, No. 336.

\item Romano E., Verde R. 2009. Clustering geostatistical data. In  Di Ciaccio A.,  Coli M., Angulo J.M.(eds). \emph{Advanced Statistical Methods for the analysis of large data-sets}. Studies in Theoretical and Applied Statistics, Springer Berlin.
\item Romano E., Balzanella A., Verde R. 2010. Clustering Spatio-functional data: a model based approach.
 \emph{Studies in Classification, Data Analysis, and Knowledge Organization}. Springer Berlin-Heidelberg, New York.
 
\item Romano E., Balzanella A., Verde R. 2010. A new regionalization method for spatially dependent functional data based on local variogram models: an application on environmental data. In: Atti delle XLV Riunione Scientifica della Societ\'{a} Italiana di Statistica Universit\'{a} degli Studi di Padova Padova. Padova, 16 -18 giugno 2010. CLEUP, ISBN/ISSN: 978 88 6129 566 7..

\item Sun, Y., and Genton, M. G. 2011. Functional boxplots, Journal of Computational and Graphical Statistics. To appear. 
  
\end{itemize}

\end{document}